\documentstyle[sprocl,amssymb,epsfig,psfig]{article}

\def\be{\begin{equation}}
\def\ee{\end{equation}}
\def\bea{\begin{eqnarray}}
\def\eea{\end{eqnarray}}

\begin{document}

\title{Quantum kinetic theory for dense Coulomb systems \\
in strong electromagnetic fields}

\author{M.~Bonitz, Th~Bornath and D.~Kremp}

\address{Universit\"at Rostock, Fachbereich Physik\\
Universit\"atsplatz 3, 18051 Rostock, Germany }

\author{H.~Haberland, M.~Schlanges and P.~Hilse}

\address{Ernst-Moritz-Arndt Universit\"{a}t Greifswald,
Institut f\"{u}r Physik, \\Domstr.~10a, 17487 Greifswald, Germany \\
E-mail: micha@elde.mpg.uni-rostock.de}

\maketitle\abstracts{A quantum kinetic theory for correlated charged--particle
systems in strong time--dependent electromagnetic fields is developed. Our
approach is based on a systematic gauge--invariant nonequilibrium Green's
functions formulation. We concentrate on the selfconsistent
treatment of dynamical screening and electromagnetic fields which is
applicable to arbitrary nonequilibrium situations. Numerical results for the
nonlinear plasma heating by the laser field and the electron--ion collision
frequency including multi--photon absorption (inverse bremsstrahlung) are
presented.}

\section{Introduction}\label{sec:intro}
With the progress in short--pulse laser technology,
high intensity electromagnetic fields are now allowing to create strongly
correlated quantum plasmas in extreme nonequilibrium conditions
with unprecedented applications \cite{perry-etal.94}.
At the same time, optical techniques for
time--resolved diagnostics are improving remarkably \cite{theobald-etal.96}
creating the need for a quantum kinetic theory of dense
nonideal plasmas in intense laser fields. A central problem is the
simultaneous account of correlations (scattering) between the particles
and of the influence of the electromagnetic field. On the other hand, such a
theory is of interest for quantum transport phenomena in solids subject to
intense THz fields, e.g. \cite{haug-jauho96}, where the common rotating wave
approximation cannot be applied.

In this paper we take advantage of the Kadanoff-Baym formalism to treat
charged particles and plasmons in a fully symmetric way. The time-depen\-dent
electromagnetic field is treated classically. Using a gauge-invariant extension
of the generalized Kadanoff-Baym ansatz, we obtain a closed equation for the
Wigner distribution which is solved for various limiting cases.

\section{Quantum kinetic equations for charged particles and longitudinal
photons}\label{sec:basis}
We consider charged particles (electrons and ions/holes/positrons etc.)
interacting via the longitudinal Coulomb force which is equivalently described
in terms of emission and absorption of longitudinal photons. The
nonequilibrium state of this system is conveniently characterized by
correlation functions of fermions
of species $a$, $g_a^{\gtrless}(1,1')$ and longitudinal photons
(screened potential) \citeup{plasmons},
$V^{s\gtrless}(1,1')$, which are defined as
averages of the respective field operator pairs \cite{qed}, where we use the
notation $1={\bf r}_1,t_1$. In particular, the density matrices of carriers
and longitudinal photons follow from the ``$^<$'' functions,
$f_a({\bf r}_1,{\bf r}_1',t_1) = -i\hbar g_a^{<}(1,1')|_{t_1=t'_1}$ and
$N({\bf r}_1,{\bf r}_1',t_1) = -i\hbar V^{s<}(1,1')|_{t_1=t'_1}$.
The transverse electromagnetic field is
given by the vector potential ${\bf A}$
and will be treated classically, i.e. it obeys Maxwell's equations
(\ref{max}). For completeness, we include also the
longitudinal field which is due to external sources, $\phi^{{\rm ext}}$,
\cite{ltf}.

The time evolution of the correlation functions is determined by the
Kada\-noff--Baym equations \cite{kadanoff-baym}
\begin{eqnarray}
&&\left[i\hbar\frac{\partial}{\partial t_1}
-\frac{1}{2m_a}\left(\frac{\hbar}{i}\nabla_1-\frac{e_a}{c}
{\bf A}(1)\right)^2 - e_a \phi^{{\rm ext}}(1)\right]g_a^{\gtrless}(1, 1')
\nonumber\\
&& \qquad \qquad\qquad- \int d{\bar {\bf r}}_1 \,
\Sigma_a^{\rm HF}(1,{\bar {\bf r}}_1t_1)
g_a^{\gtrless}({\bar{\bf r}}_1t_1,1')
\nonumber\\
&&\qquad=\int_{-\infty}^{\infty} d{\bar 1} \,
\left[\Sigma_a^R(1,{\bar 1})g_a^{\gtrless}({\bar 1}, 1') -
\Sigma_a^{\gtrless} (1,{\bar 1})\,g_a^A({\bar 1} ,1')\right],
\label{kbe}
\\
&& \Delta V_{ab}^{s\gtrless}(1, 1') =
\sum_c\int_{-\infty}^{\infty} d{\bar 1} \,
\left[\Pi_{ac}^R(1,{\bar 1})V_{cb}^{\gtrless}({\bar 1}, 1') -
\Pi_{ac}^{\gtrless} (1,{\bar 1})\,V_{cb}^{sA}({\bar 1} ,1')\right],
\label{kbei-gvs}
\\
&& \left\{\Delta - \frac{1}{c^2}\frac{\partial^2}{\partial t^2}\right\}
{\bf A}(1) = -\frac{4\pi}{c}{\bf j}(1).
\label{max}
\end{eqnarray}
The retarded and advanced quantities,
$F=g,\Sigma^{\gtrless}, \Pi$, are defined as
\begin{eqnarray}
F^{R/A}(1,1') &=&
\pm \Theta[\pm(t_1-t'_1)]\left\{F^>(1,1')-F^<(1,1')\right\},
\end{eqnarray}
whereas $g^{R/A}$ and $V^{sR/A}$ obey the equations of motion
\begin{eqnarray}
&&\left[i\hbar\frac{\partial}{\partial t_1}-\frac{1}{2m_a}
\left(\frac{\hbar}{i}\nabla_1-\frac{e_a}{c}{\bf A}(1)\right)^2
- e_a \phi^{{\rm ext}}(1)\right]
g_a^{R/A}(1,1')
\nonumber\\
&&-\int d2 \,\Sigma_a^{R/A}(1,2)\,g_a^{R/A}(2,1')\,\, =\,\,
\delta(1-1'),
\label{gra_eq}\\
&&V_{ab}^{sR/A}(1,1') =
V_{ab}(1-1')\delta(t_1-t'_1) +
\nonumber\\
&& \qquad\qquad\qquad\quad +\sum_{cd} V_{ac}(1-1')
\int d2\, \Pi_{cd}^{R/A}(1,2) \,V_{db}^{sR/A}(2,1').
\label{vra_eq}
\end{eqnarray}
($V_{ab}$ denotes the Coulomb potential between charged particles of
species ``a'' and ``b''). Eq.~(\ref{kbe}) has to be supplemented by its adjoint.
Also, we will not consider correlated initial states here, for the corresponding
generalization of the Kadanoff-Baym equations, see \cite{kremp-etal.99kbt}.

To close this system of equations, we have to specify the carrier and
longitudinal photon selfenergies $\Sigma$ and $\Pi$. The simplest ansatz
is the random phase approximation
\begin{eqnarray}
\Sigma^{\gtrless}_{a}(1,1') &=&  i\hbar \,
g_{a}^{\gtrless}(1, 1')\, V_{aa}^{s\gtrless}(1, 1'),
\nonumber\\
\Pi_{ab}^{\gtrless}(1,1') &=& -i\hbar \,
g_{a}^{\gtrless}(1, 1')\, g_{b}^{\lessgtr}(1, 1').
\label{sp-rpa}
\end{eqnarray}
Despite its simplicity, the system (\ref{kbei-gvs}) - (\ref{sp-rpa})
\cite{bonitz-etal.99cpp2} describes
remarkably complex physical processes: the evolution in space and time of
charged carriers interacting via the full dynamic Coulomb potential which in
turn evolves selfconsistently (screening build up) and may have
nonequilibrium modes (including instabilities and nonlinear phenomena).
Furthermore, the dynamics is influenced by the transverse electromagnetic
field ${\bf A}$ which contains external fields (e.g. a laser field) and
induced contributions and obeys Maxwell's equations. In this classical
treatment, ${\bf A}$ does not modify the equations for carriers and plasmons
explicitly, but only indirectly, via the particle propagators $g^{R/A}$.

Thus, a direct solution of the system (\ref{kbei-gvs}) - (\ref{sp-rpa})
would yield a tremendous amount of information. However, here
we proceed differently - we derive equations of motion for the Wigner
distributions of the charged particles $f_a({\bf k},t)$ which turn out to be
simpler. To this end we consider the correlation functions $g^{<}_a$ on the
time diagonal. To make the derivations independent on a particular
gauge, it is advantageous to make the Green's functions explicitly
gauge-invariant \cite{haug-jauho96,kremp-etal.99pre}. For the case of
a spatially homogeneous field ${\bf E}(t)$ it is convenient to choose
$\phi^{ext}=0$ and
${\bf A}(t)= - c \int^t_{-\infty} d{\bar t}\,{\bf E}({\bar t})$. Then,
the gauge-invariant Green's functions are obtained by the transform
\begin{eqnarray}
g_a({\bf k},\omega;t)=\int d\tau d{\bf r}\,\exp \left[i\,\omega \tau -
\frac{i}{\hbar}{\bf r} \left({\bf k}+
\frac{e_a}{c}\int_{t'_1}^{t_1}
\frac{d{\bar t}}{\tau}\,
{\bf A}({\bar t})\right) \right] g_a({\bf r},\tau;t),
\label{gauge-ftv}
\end{eqnarray}
where $r=r_1-r'_1$, $\tau=t_1-t'_1$ and $t=(t_1+t'_1)/2$. The resulting
quantum kinetic equation is given by \cite{kremp-etal.99pre}
\begin{eqnarray}
&&\frac{\partial}{\partial t} f_a({\bf k}_a,t)+e_a{\bf E}(t)\cdot
\nabla_{\bf k} f_a({\bf k}_a,t)=
 -2 {\rm Re}\int_{t_0}^t d{\bar t}\Big\{\Sigma_a^>g_a^< -
\Sigma_a^<g_a^>\Big\} = I_a
, \qquad
\label{f_eqgi}
\end{eqnarray}
where the full arguments of the functions on the r.h.s. are,
\begin{eqnarray}\label{sig-g}
\Sigma_a^{\gtrless}g_a^{\lessgtr}  &\equiv &
\Sigma_a^{\gtrless}\left[{\bf k}_a+{\bf K}_a^A(t,{\bar t});t,{\bar t}\,\right] \,
g_a^{\lessgtr}\left[{\bf k}_a+{\bf K}_a^A(t,{\bar t});{\bar t},t\right],
\vspace{0.6cm}
\label{ka}
\\\nonumber
\mbox{with} \quad
{\bf K}^A_a(t,t') &\equiv & \frac{e_a}{c}\int_{t'}^{t} dt'' \,
\frac{{\bf A}(t)-{\bf A}(t'')}{t-{t'}}.
\end{eqnarray}
Finally, to obtain a closed equation for the Wigner functions,
the two-time functions $\Sigma_a^{\gtrless}$ and $g_a^{\lessgtr}$ in
Eq.~(\ref{f_eqgi})  have to be expressed in terms of $f_a$. The appropriate
solution is the  generalized Kadanoff-Baym ansatz \cite{lipavski-etal.86}
which, in the case of time-dependent  fields, has the form
\cite{kremp-etal.99pre,bonitz-etal.99cpp2}
\begin{eqnarray}
\pm g_a^{\gtrless}({\bf k};t_1,t'_1) &=&  g_a^R({\bf k};t_1,t'_1) \,
f_a^{\gtrless}\left[{\bf k}-{\bf K}^A_a(t',t); t'_1\right]
\nonumber\\
&-& f_a^{\gtrless}\left[{\bf k}-{\bf K}^A_a(t,t'); t_1\right] \,
g_a^A({\bf k};t_1,t'_1),
\label{gkba_gi}
\end{eqnarray}
where the upper (lower) sign refers to $g^>$ ($g^<$), and $f^< \equiv f$
and $f^> \equiv 1-f$.

Equations (\ref{f_eqgi}) and (\ref{ka}) are valid for any choice of the
selfenergies. In particular, we can use the RPA expressions (\ref{sp-rpa}),
which leads to a collision integral $I_a$ \cite{bonitz-etal.99cpp2} which
generalizes the result of Haug and Ell \cite{haug-etal.92,haug-rpa} to the
case of an external electric field. A particularly interesting
phenomenon is that the dielectric and screening properties of the
plasma are directly modified by the electromagnetic field:
\begin{eqnarray}
\Pi^{R}_{aa}({\bf q};t,t') &=&
-\frac{i}{\hbar}\Theta(t-t')\,
e^{\frac{i}{\hbar} {\bf q}{\bf R}_a(t,t')}
\int \frac{d^3 k}{(2\pi\hbar)^3}\,
e^{-\frac{i}{\hbar}
\left(\epsilon^a_{{\bf k+q}}-\epsilon^a_{{\bf k}}\right)(t-t')
}
\nonumber\\
&\times &
\left\{ f_a\left[{\bf k}+{\bf Q}_a(t,t');t'\right]-
f_a\left[{\bf k+q}+{\bf Q}_a(t,t');t'\right]
\right\},
\label{pira-gkba}
\end{eqnarray}
where we used for $g^{R/A}$ in Eq.~(\ref{gkba_gi}) the propagators
of a free particle with energy $\epsilon$ in an electromagnetic field
\cite{johnson-etal.96}. Further, ${\bf Q}_a$ and ${\bf R}_a$ are,
respectively, the  momentum gain and displacement of a free particle in a
field $E(t)$ during  the time interval $[t',t]$ given by
\begin{equation}
{\bf Q}_a(t,t') =
- e_a\int_{t'}^{t} dt'' \, {\bf E}(t''), \quad
{\bf R}_a(t,t') =  \frac{e_a}{m_a}
\int_{t'}^{t}d{{\tilde t}}\int_{t'}^{{\tilde t}}d{\bar t}\,{\bf E}({\bar t}).
\label{QR}
\end{equation}
Obviously, the plasmon spectrum is modified in two ways: first, by the
field-dependent prefactor of the standard RPA-polarization and, second, by
the field-dependent momentum arguments of the distribution functions.

For the practical issue of plasma heating it is important to understand the
scattering  process of charged particles in the presence of the field
$E(t)$ in detail.  Analytical and numerical progress can be made if the RPA selfenergies
(\ref{sp-rpa}) are approximated by their static limit, which in
Eq.~(\ref{vra_eq}) leads to the substitution of the dynamic potential
by the statically screened Debye potential,
$V^{R/A}_s(q,t_1,t_1') \rightarrow V^{st}(q,t_1)\delta(t_1-t'_1)$.
As a result, the collision integral becomes \cite{kremp-etal.99pre}
\begin{eqnarray}
I_a({\bf k}_a,t) &=& 2 \sum\limits_{b}
\int \frac{d{\bf k}_b d{\bar {\bf k}}_a d{\bar {\bf k}}_b}{(2\pi\hbar)^9}
|V^{st}_{ab}({\bf k}_a-{\bar {\bf k}}_a)|^2
(2\pi\hbar)^3\delta ({\bf k}_{a}+{\bf k}_{b}-\bar{{\bf k}}_{a}-\bar{{\bf k}}_{b})\,
\nonumber\\
&\times & \int_{t_0}^t d {\bar t} \; \cos\left\{\frac{1}{\hbar}\left[
(\epsilon_{ab}-{\bar \epsilon}_{ab})(t-\bar{t})-
({\bf k}_a-{\bar {\bf k}}_a){\bf R}_{ab}(t,{\bar t})
\right]\right\}
\nonumber\\
&&\times
\qquad\left\{{\bar f}_a {\bar f}_b\,[1-f_a] \,[1-f_b] -
f_a f_b \,[1-{\bar f}_a] \, [1-{\bar f}_b]\right\}\big|_{{\bar t}}\:,
\qquad
\label{f_eqfin}
\end{eqnarray}
where we denoted $\epsilon_{ab} \equiv \epsilon_{a}+\epsilon_{b}$,
$\epsilon_{a}\equiv p^2_a/2m_a$,
$f_a \equiv f(K_a,{\bar t})$, ${\bar f}_a \equiv f({\bar K}_a,{\bar t})$,
 ${\bf K}_a \equiv {\bf k}_a+{\bf Q}_a$ and
 ${\bf R}_{ab} \equiv {\bf R}_{a}-{\bf R}_{b}$. Although plasmon and screening
dynamics are no longer included, Eq.~(\ref{f_eqgi}) with the collision term
(\ref{f_eqfin}) still contains important physics reflecting the influence of
the electromagnetic field: (I) field-induced change of the arguments of the
distribution functions, i.e. time-dependent generalization of the
intra-collisional field effect; (II) modification of the energy balance in
two-particle scattering [argument of the cosine in Eq.~(\ref{f_eqfin})] -
a strong field may essentially modify the energy broadening in the
electron-ion integral; (III) nonlinear (exponential)
dependence of the collision integral on the field strength - this leads to the
generation of higher field harmonics in the scattering processes. Furthermore,
it gives rise to scattering processes which involve emission (absorption) of
photons, i.e. (inverse) bremsstrahlung. Indeed, it is straightforward to show
\cite{kremp-etal.99pre} that, for an electric field $E(t)=E_0 \rm{cos}\Omega
t$ transport quantities computed from (\ref{f_eqfin}) will contain
contributions proportional to $J_n^2(z)
\delta[\epsilon_{ab}-{\bar \epsilon}_{ab}+{\bf q}{\bf w}_{ab}(t)-n\hbar \Omega]$,
($-\infty < n < \infty$),
where the amplitude of an n-photon process is given by the Bessel functions
$J_n$. The argument $z$ of $J_n$ is determined by the field strength and frequency,
$z={\bf q}[{\bf v}^0_{a}-{\bf v}^0_{b}]/\hbar\Omega$, where
$v_a^0=e_a E_0/m_a\Omega$ and
${\bf w}_{ab}=[{\bf v}^0_{a}-{\bf v}^0_{b}]\sin\Omega t$.

\section{Numerical results}\label{sec:res}
The kinetic equation (\ref{f_eqgi}) with the collision integral
(\ref{f_eqfin}) is a convenient starting point for numerical investigations.
We underline that it fully contains quantum effects, therefore, in contrast
to classical equations, no cutting procedures at large momenta are
required. Before presenting results of
direct numerical solutions, we outline the idea for an approximate
treatment \cite{bonitz-etal.99cpp2,bornath-etal.99lpb}.

If in Eq.~(\ref{f_eqgi}) the collision term is much smaller than the
field term on the l.h.s., one can use an ansatz proposed by Silin
\cite{silin64},
$f_a=f_a^0+f_a^1$, where $f_a^0$ obeys the collisionless equation, with
the solution
$f^0_a({\bf k},t) = f_{a0}\left[{\bf k}+\frac{e_a}{c}{\bf A}(t)\right]$,
and the correction $f^1_a$ follows from Eq.~(\ref{f_eqgi}) with
$I_a[f_a] \rightarrow I_a[f^0_a]$. Further simplifications are possible
if the distributions are Maxwellians. Then, most of the integrations
in the collision term can be performed, and analytical results for
field-dependent transport quantities can be derived such as the
conductivity, Fig.~1, or the cycle-averaged energy gain per particle,
$\langle\, {\bf j} \cdot {\bf E}\,\rangle =
\frac{1}{T}\int_{t-T}^{t} dt'\,
{\bf j}(t')\cdot {\bf E}(t')$, for which one obtains
\cite{bornath-etal.99lpb}
\begin{eqnarray}\label{energy}
\langle\, {\bf j} \cdot {\bf E}\,\rangle
&=&\frac{8\sqrt{2\pi}Z^2 e^4 n_en_i
\sqrt{m_e}}{(k_BT)^{3/2}}\,\Omega^2\,
\sum_{n=1}^{\infty}\,n^2\,\int\limits_0^\infty
dk\,\frac{k}{(k^2+k_D^2)^2}\,
\exp{\left\{-\frac{n^2m_e\Omega^2}{2k_BT\,k^2}\right\}}\,
\nonumber\,\\[1ex]
&&\times
\exp{\left\{-\frac{\hbar^2 k^2}{8m_ek_BT}\right\}}\,\,
\frac{
\sinh{\frac{n\hbar\Omega}{2k_BT}}
}
{\frac{n\hbar\Omega}{2k_BT}}\,\,
\int_0^1 dz\,J_n^2\left(\frac{eE_0k}{m_e\Omega^2}\,z\right),
\end{eqnarray}
with the Debye screening wave number $k_D^2=\lambda_D^{-2}=4\pi n_e e^2/k_BT$.
We underline that quantum effects occur in (\ref{energy}) in two places:
first, the exponential function in the second line
automatically ensures the convergence of the integral for large
$k$ and, second, the  Bose-Einstein statistics of
photons is reflected by the factor with the $\sinh$ function.
For a classical plasma, both factors approach $1$, and one recovers the
result of Klimontovich \cite{klimontovich80}.
Fig.~1 shows results for the conductivity computed using the Silin
ansatz with the statically screened and the dynamically screened
e-i collision integral. Also, we show a fit formula of Silin which holds
for classical plasmas, but fails at low temperatures.

Finally, we present results of direct numerical integration of
the quantum kinetic equation (\ref{f_eqgi}) with the collision term
(\ref{f_eqfin}). This has the advantage that no assumption on weak
collisions or shape of the distributions has to be made. The
heating of the electrons for three different densities is shown
in Fig.~2, and Fig.~3 shows the evolution of the electron distribution
during the first laser cycle. One clearly sees the anisotropy of
the distribution and its broadening due to absorption of laser energy.
A detailed analysis shows that, in fact, multi photon absorption
(inverse bremsstrahlung) occurs which will be discussed elsewhere.
\begin{figure}[paul_s]
\vspace*{-0.3cm}
\centerline{
\psfig{file=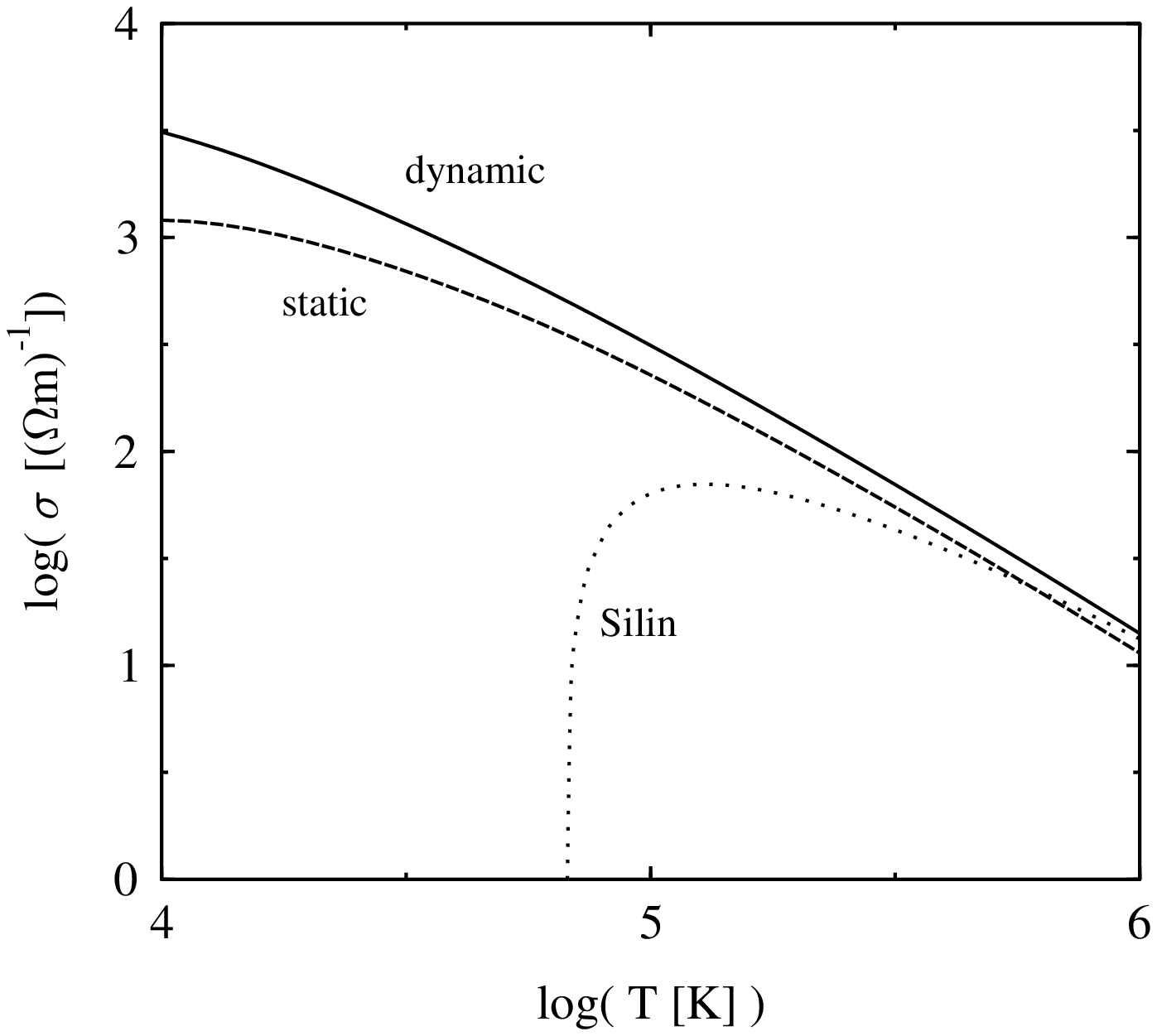,height=6.cm,width=9cm}}
\vspace{0.02cm}
\caption[]{\label{s} Conductivity of a hydrogen plasma with density
$n=5 \cdot 10^{21}cm^{-3}$ vs. temperature
in a strong high frequency laser field with $\lambda=248.5$nm,
$\Omega/\omega_{pl}=1.9$, $v_0/v_{th}=5$.
}
\vspace*{-1.3cm}
\end{figure}

\begin{figure}[hagen_e]
\vspace*{-1cm}
\centerline{
\psfig{file=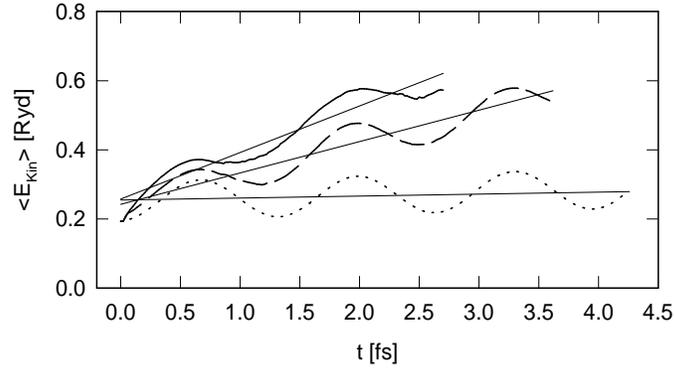,height=5cm,width=9cm}}
\vspace{0.02cm}
\caption[]{\label{ekin} Electron heating due to Coulomb collisions in a
strong laser field for different densities: $n_e=10^{22}cm^{-3}$ (full line),
$n_e=10^{23}cm^{-3}$ (dashes) and $n_e=10^{24}cm^{-3}$ (dots). Remaining
parameters same as in Fig.~3. (The straight lines are a guide for the
eye.) }
\vspace*{-0.3cm}
\end{figure}

\begin{figure}[hagen_f]
\vspace*{-0.3cm}
\centerline{
\psfig{file=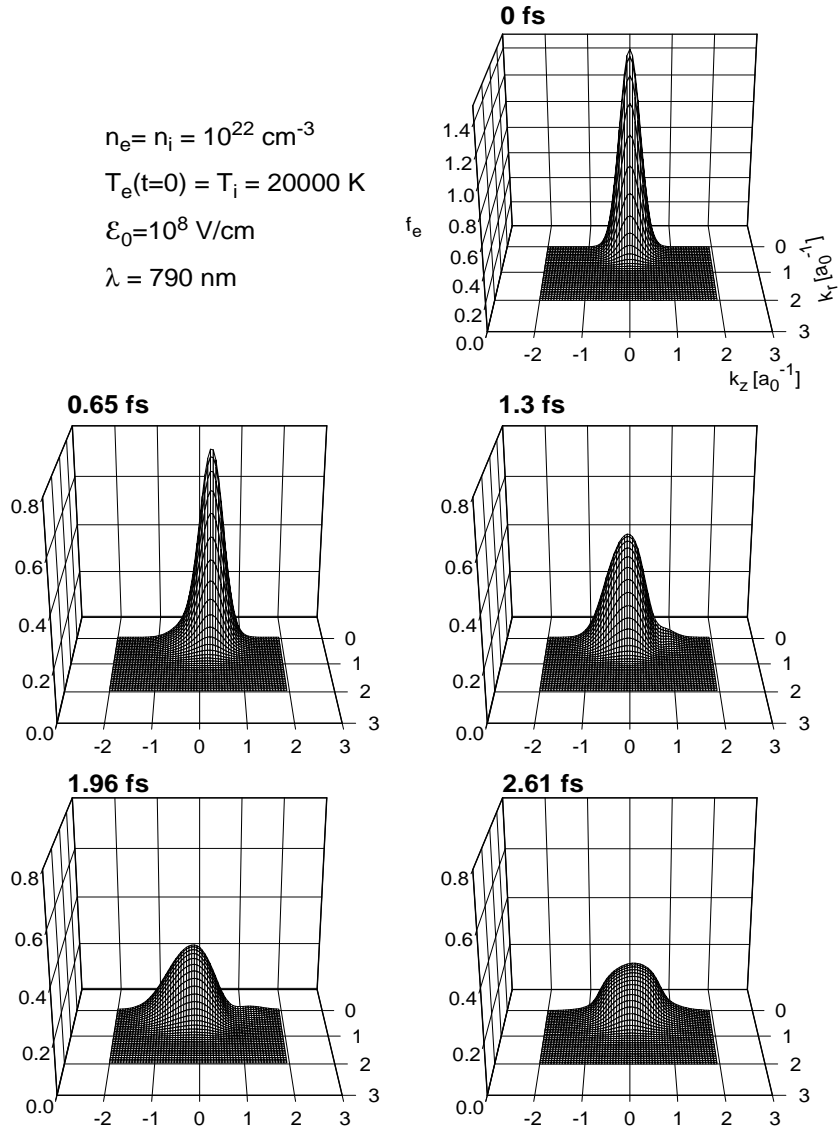,height=15cm,width=11cm}}
\vspace{0.02cm}
\caption[]{\label{f} Evolution of the electron distribution during the
first laser period. $E(t=0) = 10^8 V/cm$, figures correspond to subsequent
quarter periods. Parameters are for hydrogen, see inset.
}
\vspace*{-0.3cm}
\end{figure}

\section*{Acknowledgments}
This work is supported by the Deutsche Forschungsgemeinschaft (Schwerpunkt
``Laserfelder'') and by the European Commission through the TMR Network
SILASI. M.~B. acknowledges discussions with Don DuBois.


\begin{thebibliography}{99}
\bibitem{plasmons} Notice that the term ``longitudinal photons'' is more
general than the commonly used term ``plasmons'' as it includes besides
bound electron-hole excitations (plasmons) also scattering states. In nonequilibrium
(in terms of two-time functions) there is no clear separation between
bound and scattering contributions.
\end{thebibliography}
\end{document}